\documentclass[11pt]{article}
\usepackage[a4paper,margin=1in]{geometry}

\usepackage[dvipdfmx]{graphicx}
\usepackage{bxtexlogo}
\usepackage{amsmath}
\usepackage{amssymb}
\usepackage{amsthm}
\usepackage{mathtools}
\usepackage{graphicx}
\usepackage{braket}
\usepackage{bm}
\usepackage{bbm}
\usepackage{comment}
\usepackage{float}
\usepackage{ascmac}
\usepackage{url}
\usepackage{cite}
\usepackage[colorlinks=true, allcolors=blue]{hyperref}

\usepackage{mathrsfs}

\DeclareMathOperator{\Tr}{Tr}

\newtheorem{dfn}{Definition}
\newtheorem{prop}[dfn]{Proposition}

\newtheorem{cor}[dfn]{Corollary}
\newtheorem{rem}[dfn]{Remark}

\def\Bcal{{\cal B}}

\def\Dcal{{\cal D}}
\def\Ecal{{\cal E}}
\def\Fcal{{\cal F}}
\def\Hcal{{\cal H}}
\def\Ical{{\cal I}}

\def\Ocal{{\cal O}}

\def\Scal{{\cal S}}
\def\Ucal{{\cal U}}
\def\Lsf{{\mathsf{L}}}
\def\Ssf{{\mathsf{S}}}

\def\Usf{{\mathsf{U}}}
\def\dim{\hbox{dim}}
\def\one{{\mathbbm{1}}}

\newcommand{\ketbra}[2]{|{#1}\rangle\!\langle{#2}|}

\renewcommand{\ge}{\geqslant}
\renewcommand{\geq}{\geqslant}
\renewcommand{\le}{\leqslant}

\allowdisplaybreaks

\usepackage[affil-it]{authblk}
\newcommand{\email}[1]{\href{mailto:#1}{#1}}

\title{Virtual phase-covariant quantum broadcasting for qubits}

\author[ ]{Reiji Okada\thanks{\email{reiji.okada@nagoya-u.jp}}}

\author[ ]{Francesco Buscemi\thanks{\email{buscemi@nagoya-u.jp}}}

\affil[ ]{Department of Mathematical Informatics, Nagoya University, Japan}

\date{\vspace{-5ex}}

\begin{document}
\maketitle
\begin{abstract}
Virtual maps allow the simulation of quantum operations by combining physical processes with classical post-processing. Recent work on virtual unitary covariant broadcasting has shown, however, that such maps remain impractical for observable estimation tasks due to poor sample efficiency. Here we investigate whether relaxing the symmetry requirements can improve operational performance, focusing on virtual phase-covariant quantum broadcasting for qubits. We show that imposing phase-covariance, flip covariance, permutation invariance, and classical consistency fully determines the structure of the broadcasting map. Within this family, we identify the unique map that minimizes the simulation cost, and we prove that both the simulation cost and the distance to the closest CPTP map are strictly smaller than in the unitary covariant setting. We also demonstrate that the closest physical map is the optimal phase-covariant cloning channel, mirroring the relation between unitary covariant broadcasting and universal cloning. Despite these improvements, the resulting virtual broadcasting map remains sample-inefficient and is therefore still operationally impractical.
\end{abstract}

\section{Introduction}

It is a trivial fact that, in classical information theory, information can be copied. In quantum information theory, by contrast, fundamental constraints prevent such copying. The no-cloning theorem~\cite{wootters1982single,DIEKS1982271} states that no completely positive trace-preserving (\texttt{CPTP}) map can create a perfect copy of an unknown quantum state. The no-broadcasting theorem~\cite{barnum1996noncommuting} strengthens this insight: if the set of states to be broadcast contains at least one pair of noncommuting states, then no \texttt{CPTP} map can output systems whose reduced states all coincide with the input. Broadcasting is less demanding than cloning because it allows the outputs to be entangled, yet even this weaker task cannot be achieved physically.

To work around this limitation, recent studies have explored \textit{virtual maps}~\cite{parzygnat2024virtual,yao2024optimal,zheng2025experimental,xiao2025noPRL,xiao2025noPRR,yao2025quantifying}. A virtual map is a Hermitian-preserving trace-preserving (\texttt{HPTP}) map that need not correspond to any physical process. Instead, it reproduces the expectation values that would result from broadcasting by combining physical operations with classical post-processing~\cite{buscemi2013direct,buscemi2014universal,rossini2023single}. Parzygnat \textit{et al.}~\cite{parzygnat2024virtual} showed that unitary covariance, permutation invariance, and classical consistency single out a unique virtual broadcasting map, known as the \textit{canonical broadcasting map}. Although not physically implementable, this map has found applications in the evaluation of two-point correlation functions~\cite{buscemi2013direct,buscemi2014universal}, error mitigation~\cite{temme2017error,jiang2021physical,takagi2022fundamental}, quantum states over time~\cite{fullwood2022quantum,parzygnat2023time,lie2023uniqueness}, and the simulation of non-Markovian dynamics~\cite{rossini2023single}. Indeed, the $1\to 2$ virtual broadcasting map was first developed to reproduce two-point correlation functions~\cite{buscemi2013direct,buscemi2014universal}, and this remains one of its most natural applications. However, if one insists on using it merely to mimic access to two independent copies of the input state, the question arises whether its simulation cost is justified when compared with the trivial redistribution of the available copies.

This issue is central to the operational meaning of virtual maps. Because they are implemented through sampling and post-processing, they typically require multiple copies of the input state. This leads to a natural question: when estimating expectation values on two outputs, is it more efficient to use a virtual broadcasting map or simply distribute the available input copies to two parties? Xiao \textit{et al.}~\cite{xiao2025noPRL,xiao2025noPRR} showed that direct distribution is strictly more efficient, which reveals the large simulation cost of the canonical map. This motivates the search for alternative virtual maps with reduced cost, obtained by relaxing some of the assumptions used in the unitary covariant construction.

A similar situation occurs in quantum cloning, where symmetry restrictions often improve performance. Cloning channels are frequently designed to satisfy group covariance, meaning that they act uniformly on all states invariant under a chosen symmetry group~\cite{werner1998optimal,d2001optimal,d2003optimal,buscemi2005economical}. Two prominent examples are \textit{unitary} (or universal) covariance, which imposes uniform fidelity for all pure states, and \textit{phase-covariance}, which requires uniform fidelity only for states on a restricted subset, such as the equator of the Bloch sphere in the qubit case. These restricted settings typically enable higher cloning fidelities and more economical implementations. By analogy, focusing on a reduced class of input states in virtual broadcasting may lower the simulation cost.

In this work, we investigate virtual broadcasting for qubits under phase-covariance. Our approach replaces the unitary covariance used in~\cite{parzygnat2024virtual,xiao2025noPRL,xiao2025noPRR} with phase-covariance and supplements it with bit-flip covariance, permutation invariance, and classical consistency. We show that these conditions already determine the structure of the virtual map on phase-covariant input states and, importantly, that they imply the broadcasting condition without assuming it explicitly. We identify the unique virtual map of minimal simulation cost and show that both this cost and the distance to the nearest physical map are smaller than in the unitary covariant case. We also show that the closest physical map is the optimal phase-covariant cloning channel. Nevertheless, despite these improvements, the resulting virtual broadcasting map remains less sample-efficient than simply distributing input copies directly.

The paper is organized as follows. In Sec.~\ref{sec: Preliminaries}, we review quantum broadcasting and introduce the symmetry and consistency conditions used in our analysis. In Sec.~\ref{sec: Virtual phase-covariant quantum broadcasting for qubits}, we derive and characterize the virtual phase-covariant broadcasting map, evaluate its simulation cost, and compare it with the optimal physical map. In Sec.~\ref{sec: Conclusion}, we summarize the results and discuss open problems.

\section{Preliminaries}\label{sec: Preliminaries}


A two-dimensional quantum system is associated with a Hilbert space $\Hcal\cong\mathbb{C}^2$. We denote the set of linear operators on $\Hcal$ as $\Lsf(\Hcal)$, the set of density operators as $\Ssf(\Hcal)$, and the set of unitary operators as $\Usf(\Hcal)$. We denote the identity operator on $\Hcal$ as $\one$ and the identity channel on $\Lsf(\Hcal)$ as $\Ical$. Throughout the paper, $\Bcal$ denotes a linear map from $\Lsf(\Hcal)$ to $\Lsf(\Hcal\otimes\Hcal)$. We define the Choi operator of $\Bcal$ as $C_\Bcal\coloneq\sum_{i,j}\Bcal(\ketbra{i}{j})\otimes\ketbra{i}{j}$, where $\{\ket{i}\}_i$ is a fixed orthonormal basis.

Quantum broadcasting refers to an operation that generates an output state whose reduced states on each subsystem coincide with the input state. A map $\Bcal:\Lsf(\Hcal) \to \Lsf(\Hcal\otimes\Hcal)$ is called a $1\to 2$ $\mathscr{S}$-\textit{broadcasting map} if
\begin{align}\label{eq: BC}
\Tr_1\Bcal(\rho)=\Tr_2\Bcal(\rho)=\rho
\end{align}
holds for every state $\rho$ in the subset $\mathscr{S}\subseteq
\Ssf(\Hcal)$. We refer to condition \eqref{eq: BC} as the $\mathscr{S}$-\textit{broadcasting condition} ($\mathscr{S}$-\texttt{BROAD}).

The no-broadcasting theorem~\cite{barnum1996noncommuting} states that there is no \texttt{CPTP} map that satisfies $\mathscr{S}$-\texttt{BROAD} whenever $\mathscr{S}$ contains at least on pair of noncommuting states. However, it has been shown that a unique \texttt{HPTP} map that satisfies $\Ssf(\Hcal)$-\texttt{BROAD}, namely, broadcasting \textit{everywhere} on $\Ssf(\Hcal)$, is singled out by the following three assumptions~\cite{parzygnat2024virtual,xiao2025noPRL,xiao2025noPRR}:
\begin{enumerate}

\item the first is that $\Bcal$ satisfies \textit{unitary covariance} (\texttt{UNITARY}), expressed as
\begin{align}\label{eq: UC}
(\Ucal\otimes\Ucal)\circ\Bcal=\Bcal\circ\Ucal,
\end{align}
where $\Ucal:\Lsf(\Hcal)\to\Lsf(\Hcal)$ is defined by $\Ucal(\cdot)\coloneq U\cdot U^\dagger$ for any arbitrary but fixed $U\in\Usf(\Hcal)$;

\item the second assumption is \textit{permutation invariance} (\texttt{PERM}), expressed as
\begin{align}\label{eq: PI}
\Scal\circ\Bcal=\Bcal,
\end{align}
where $\Scal:\Lsf(\Hcal\otimes\Hcal)\to\Lsf(\Hcal\otimes\Hcal)$ is defined by $\Scal(\cdot)\coloneq S\cdot S$ for the swap operator $S:\Hcal^{\otimes 2}\to \Hcal^{\otimes 2}$ defined by $S(\ket{\phi}\otimes\ket{\psi})=\ket{\psi}\otimes\ket{\phi}$, for any $\ket{\phi},\ket{\psi}\in\Hcal$;

\item the third assumption requires that, when both the input and output systems of $\Bcal$ undergo decoherence, its action coincides with the classical broadcasting map. The classical broadcasting map $\Bcal_\mathrm{cl}:\Lsf(\Hcal)\to\Lsf(\Hcal\otimes\Hcal)$ is the unique map defined by 
\begin{align}
\Bcal_\mathrm{cl}(\ketbra{i}{j})\coloneq\delta_{ij}\ketbra{i}{i}\otimes\ketbra{i}{i}\quad\forall i,j\;,
\end{align}
for a fixed orthonormal basis $\{\ket{i}\}_i$. We say that $\Bcal$ satisfies \textit{classical consistency} (\texttt{CLASSIC}) if
\begin{align}\label{eq: CC}
(\Dcal\otimes\Dcal)\circ\Bcal\circ\Dcal=\Bcal_\mathrm{cl},
\end{align}
where $\Dcal:\Lsf(\Hcal)\to\Lsf(\Hcal)$ is the decoherence map defined by $\Dcal(\rho)\coloneq\sum_i\bra{i}\rho\ket{i}\ketbra{i}{i}$. If a broadcasting $\Bcal$ satisfies \texttt{UNITARY} and \texttt{CLASSIC} for a certain basis,  then it automatically satisfies \texttt{CLASSIC} also for any other orthonormal basis~\cite{parzygnat2024virtual}.

\end{enumerate}

The assumption of \textit{phase-covariance} (\texttt{PHASE}) used in this work is a relaxation of \texttt{UNITARY} covariance, obtained by restricting attention to a subset of $\Ssf(\Hcal)$. We focus on
\begin{align}
    \Ssf'(\Hcal)\coloneq&\ \{\rho\mid \rho\in\Ssf(\Hcal),\ \Tr[\rho\sigma_z]=0\} \\
    =&\ \left\{\tfrac{1}{2}(\one+r\cos\varphi\,\sigma_x+r\sin\varphi\,\sigma_y)\ \middle|\ 0\le r\le1,\ 0\le\varphi<2\pi\right\},
\end{align}
namely the equatorial states of the Bloch sphere. The relevant symmetry group is the set of diagonal unitaries
\begin{align}
    \Usf'(\Hcal)\coloneq\{U_\varphi\mid 0\le\varphi<2\pi\},
\end{align}
where $U_\varphi\coloneq\ket{0}\bra{0}+e^{i\varphi}\ket{1}\bra{1}$; global phases are immaterial. Condition \texttt{PHASE} demands covariance only with respect to $\Usf'(\Hcal)$:
\begin{align}
    (\Ucal'\otimes\Ucal')\circ\Bcal=\Bcal\circ\Ucal',
\end{align}
for any $\Ucal'(\cdot)\coloneq U_\varphi\cdot U_\varphi^\dag$ with $U_\varphi\in\Usf'(\Hcal)$. In this setting, both \texttt{BROAD} and \texttt{CLASSIC} are required only on states in $\Ssf'(\Hcal)$. In particular, in the phase-covariant setting it is natural to impose \texttt{CLASSIC} with respect to equatorial bases; indeed, in Sec.~\ref{sec: Virtual phase-covariant quantum broadcasting for qubits} we formulate \texttt{CLASSIC} starting from the $\{\ket{+},\ket{-}\}$ basis, where $\ket{\pm}\coloneq\frac{1}{\sqrt{2}}(\ket{0}\pm\ket{1})$.

Because $\Ssf'(\Hcal)$ is closed under conjugation by $\sigma_x$, it is natural to supplement \texttt{PHASE} with covariance under a bit-flip about the $x$–axis. Together, these symmetries imply covariance under all reflections of the equatorial plane. A map $\Bcal$ satisfies \textit{flip covariance} (\texttt{FLIP}) if
\begin{align}\label{eq: FC}
    (X\otimes X)\circ\Bcal=\Bcal\circ X,
\end{align}
where $X(\cdot)\coloneq\sigma_x\cdot\sigma_x$.

\subsection{Sample efficiency}

Although broadcasting cannot be physically realized, its expectation values can be simulated through virtual operations, namely \texttt{HPTP} maps. To assess whether such simulations are operationally meaningful, we use the notion of \textit{sample efficiency} (\texttt{SAMPLE})~\cite{xiao2025noPRL,xiao2025noPRR}. \texttt{SAMPLE} compares the number of input copies required by a virtual operation with that of the trivial strategy where Alice directly distributes copies of the unknown input state \(\rho\) to each party.

Consider the following task. Alice holds many copies of an unknown state \(\rho\). She sends systems to Bob and Claire, who locally measure observables \(\Ocal_1\) and \(\Ocal_2\), respectively. To estimate the corresponding expectation values with desired accuracy, Bob requires \(n_1\) copies of \(\rho\), and Claire requires \(n_2\) copies. The trivial solution is for Alice to distribute these \(n_1+n_2\) copies directly.

Virtual operations offer another possibility. A virtual broadcasting map allows Bob and Claire to reproduce the correct expectation values, but the required number of input copies now depends on how \(\rho\) is processed through the map and on the classical post-processing used in the simulation. If a virtual broadcasting map enables Bob and Claire to achieve their respective accuracies while Alice uses fewer than \(n_1+n_2\) copies of \(\rho\), the map is regarded as practical. Otherwise, the trivial distribution strategy is already optimal. \texttt{SAMPLE} therefore demands that the simulation cost of a virtual broadcasting map be strictly smaller than that of the trivial method.

We now formalize this requirement. Bob and Claire must each pass an \(\epsilon_i\)–\(\delta_i\) test: the probability that their estimation error exceeds \(\epsilon_i\) is at most \(\delta_i\). By Hoeffding's inequality, the number of copies needed is bounded below by
\begin{align}\label{eq: lower bound of n}
n_i \ge \frac{c^2}{2\epsilon_i^2}\ln\frac{2}{\delta_i} \qquad (i\in\{1,2\}),
\end{align}
where \(c\) is the range of possible measurement outcomes.

Next, consider a \texttt{HPTP} map \(\Bcal\) decomposed as \(\Bcal = a\Ecal - b\Fcal\), where \(\Ecal\) and \(\Fcal\) are \texttt{CPTP} maps and \(a,b\ge 0\). Then
\begin{align}
\Tr[\Ocal_1\Tr_2\Bcal(\rho)]
&= a\Tr[\Ocal_1\Tr_2\Ecal(\rho)]-b\Tr[\Ocal_1\Tr_2\Fcal(\rho)]\\
&=(a+b)\!\left(\frac{a}{a+b}\Tr[\Ocal_1\Tr_2\Ecal(\rho)]-\frac{b}{a+b}\Tr[\Ocal_1\Tr_2\Fcal(\rho)]\right).
\end{align}
Thus, to estimate \(\Tr[\Ocal_1\Tr_2\Bcal(\rho)]\), one samples \(\Ecal\) with probability \(a/(a+b)\) and \(\Fcal\) with probability \(b/(a+b)\), computes the difference of their expectation values, and rescales by \(a+b\). The rescaling amplifies the estimation error by the same factor. Combining this with \eqref{eq: lower bound of n}, simulating \(\Bcal\) requires \((a+b)^2 n_{\mathrm Q}\) copies of \(\rho\), where \(n_{\mathrm Q} := \max\{n_1,n_2\}\).

We say that \(\Bcal\) satisfies \texttt{SAMPLE} when
\[
n_1+n_2 > \min (a+b)^2 n_{\mathrm Q},
\]
where the minimum is taken over all decompositions of \(\Bcal\) into two \texttt{CPTP} maps.

To determine whether a \texttt{HPTP} map \(\Bcal\) satisfies \texttt{SAMPLE}, one must therefore identify the decomposition minimizing \(a+b\). Following Ref.~\cite{buscemi2013direct} and the framework of Ref.~\cite{regula2021operational}, this optimization is expressed through the base norm associated with completely positive trace-non-increasing (\texttt{CPTNI}) maps:
\begin{align}\label{eq: base norm}
\|\Bcal\|_\blacklozenge \coloneqq \min\{a+b \mid \Bcal = a\Ecal - b\Fcal,\ \Ecal,\Fcal:\text{\texttt{CPTNI}}\}.
\end{align}
The use of \texttt{CPTNI} maps is needed only when \(\Bcal\) is not trace-preserving. When \(\Bcal\) is \texttt{TP}, Ref.~\cite{regula2021operational} shows that the minimization can be restricted to \texttt{CPTP} maps:
\begin{align}
\|\Bcal\|_\blacklozenge=\min\{a+b \mid \Bcal=a\Ecal-b\Fcal,\ a,b\ge 0,\ \Ecal,\Fcal:\text{\texttt{CPTP}}\}.
\end{align}
Since this work considers only trace-preserving maps, definition \eqref{eq: base norm} is appropriate. We refer to \(\|\Bcal\|_\blacklozenge\) as the \textit{simulation cost} of \(\Bcal\).

Finally, the following bound relates the base norm to the trace norm of the Choi operator \(C_\Bcal\)~\cite{regula2021operational}:
\begin{align}\label{eq: bound of simulation cost}
\frac{1}{d}\|C_\Bcal\|_1 \le \|\Bcal\|_\blacklozenge \le \|C_\Bcal\|_1,
\end{align}
where \(d = \dim(\Hcal)\) and \(\|\cdot\|_1\) denotes the trace norm. This inequality provides tractable estimates for simulation cost and therefore for sample efficiency.

\section{Virtual phase-covariant quantum broadcasting for qubits}\label{sec: Virtual phase-covariant quantum broadcasting for qubits}

The group covariance of a map is expressed as the invariance of its Choi operator under the corresponding group action~\cite{d2001optimal}. When $\Bcal$ satisfies \texttt{PHASE}, the Choi operator $C_\Bcal$ obeys
\begin{align}
(U_\varphi\otimes U_\varphi\otimes U_\varphi^*)\ C_\Bcal\ (U_\varphi\otimes U_\varphi\otimes U_\varphi^*)^\dagger
= C_\Bcal \qquad \forall\, U_\varphi\in\Usf'(\Hcal),
\end{align}
where $U_\varphi^*$ denotes the complex conjugate of $U_\varphi$. Since
\begin{align}
(U_\varphi\otimes U_\varphi\otimes U_\varphi^*)\ket{jkl}
= e^{i(j+k-l)\varphi}\ket{jkl} \qquad (j,k,l\in\{0,1\}),
\end{align}
the joint action of $U_\varphi$, the swap $S$, and the bit-flip $\sigma_x$ decomposes the space $\Hcal^{\otimes3}$ into the invariant subspaces listed in Table~\ref{tab: decomposition}. By Schur’s lemma, the Choi operator takes the block form
\[
C_\Bcal=\bigoplus_{j=1}^6 C_j,
\]
where each block $C_j$ acts on the corresponding subspace $\Hcal_j$.

\begin{table}[tbh]
\centering
\begin{tabular}{ccccc}
\hline
Subspace     & Unnormalized Basis                & $U_\varphi$ & $S$ & $\sigma_x$ \\ \hline
$\Hcal_1$ & $\ket{000},\ \ket{011}+\ket{101}$ & 0        & 1   & $\Hcal_2$  \\
$\Hcal_2$ & $\ket{111},\ \ket{100}+\ket{010}$ & 1        & 1   & $\Hcal_1$  \\
$\Hcal_3$ & $\ket{011}-\ket{101}$             & 0        & -1  & $\Hcal_4$  \\
$\Hcal_4$ & $\ket{100}-\ket{010}$             & 1        & -1  & $\Hcal_3$  \\
$\Hcal_5$ & $\ket{001}$                       & -1       & 1   & $\Hcal_6$  \\
$\Hcal_6$ & $\ket{110}$                       & 2        & 1   & $\Hcal_5$  \\ \hline
\end{tabular}
\caption{Decomposition of $\Hcal^{\otimes3}$ into equivalence classes and invariant subspaces, under the combined action of $U_\varphi$, the swap $S$, and the bit-flip $\sigma_x$. The first column labels the inequivalent representations, and the second column lists irreducible representations (all one-dimensional) within the same equivalence class. The integer $n$ in the third column specifies the character of the action of $U_\varphi\otimes U_\varphi\otimes U_\varphi^*$ on that subspace, which acts as multiplication by $e^{in\varphi}$. The sign in the fourth column records the eigenvalue of the operator $S_{12}\otimes\one_3$. The fifth column indicates the partner class obtained from the action of $\sigma_x\otimes\sigma_x\otimes\sigma_x^*$. Note that the superscript $*$ denotes complex conjugation: since $\sigma_x$ is real, here it plays no role, but we keep it for clarity.}
\label{tab: decomposition}
\end{table}

The condition \texttt{FLIP} imposes an additional restriction. The operator $\sigma_x\otimes\sigma_x\otimes\sigma_x^*$ maps $\Hcal_j$ onto $\Hcal_{j+1}$ for $j=1,3,5$. Because $C_\Bcal$ is invariant under this action, the two blocks on each such pair of subspaces must coincide. In other words, $C_j = C_{j+1}$ as operators for $j=1,3,5$. We therefore use the same scalar parameters to describe each pair. We denote the parameters for $C_1,\ C_3$, and $C_5$ by $\{c_1,c_2,c_3,c_4\},\ c_5$, and $c_6$, respectively, with $c_1,\ldots,c_6\in\mathbb{C}$.

In the computational basis, the Choi operator of a map $\Bcal$ satisfying \texttt{PHASE}, \texttt{FLIP}, and \texttt{PERM} can be written as
\begin{align}\label{eq: PCPIFC}
C_\Bcal=
\begin{bmatrix}
c_1 & 0 & 0 & c_2 & 0 & c_2 & 0 & 0 \\
0 & c_6 & 0 & 0 & 0 & 0 & 0 & 0 \\
0 & 0 & c_4+c_5 & 0 & c_4-c_5 & 0 & 0 & c_3 \\
c_3 & 0 & 0 & c_5+c_5 & 0 & c_4-c_5 & 0 & 0 \\
0 & 0 & c_4-c_5 & 0 & c_4+c_5 & 0 & 0 & c_3 \\
c_3 & 0 & 0 & c_4-c_5 & 0 & c_4+c_5 & 0 & 0 \\
0 & 0 & 0 & 0 & 0 & 0 & c_6 & 0 \\
0 & 0 & c_2 & 0 & c_2 & 0 & 0 & c_1 \\
\end{bmatrix}.
\end{align}

Since here we consider qubit states lying on the $xy$-plane (equatorial states), the decoherence map $\Dcal$ and the classical broadcasting map $\Bcal_\mathrm{cl}$ with respect to which \texttt{CLASSIC} is defined are taken with respect to the basis $\{\ket{+},\ket{-}\}$. Using the reverse-Choi relation $\Bcal(\rho)=\Tr_3[\one_{12}\otimes\rho^T_3\ C_\Bcal]$, we obtain
\begin{align}\label{eq: bCC}
(\Dcal\otimes\Dcal)\circ\Bcal(\ketbra{+}{+})=&\ \left(\frac{c_1+c_6}{4}+\frac{c_2+c_3}{2}+c_4\right)\ketbra{++}{++}\nonumber\\
+&\ \left(\frac{c_1+c_6}{4}+c_5\right)\big(\ketbra{+-}{+-}+\ketbra{-+}{-+}\big)\nonumber\\
+&\ \left(\frac{c_1+c_6}{4}-\frac{c_2+c_3}{2}-c_4\right)\ketbra{--}{--}.
\end{align}
Then, as a consequence of linearity and \texttt{PHASE}, a necessary and sufficient condition for $\Bcal$ to satisfy \texttt{CLASSIC}, is that \eqref{eq: bCC} equals $\ketbra{++}{++}$, which is in turn equivalent to the following set of simultaneous linear conditions:
\begin{align}
\begin{cases}\label{eq: condition for CC}
c_3=1-c_2\;,\\
c_5=c_4-\frac{1}{2}\;,\\
c_6=2-c_1-4c_4\;.
\end{cases}
\end{align}
Note that, since we work with qubits and \texttt{PHASE} covariance, knowing the action of $\Bcal$ on $\ketbra{+}{+}$ already determines its action on $\ketbra{-}{-}$, because $\ket{-}=\sigma_z\ket{+}$ and phase-covariance fixes the behavior of $\Bcal$ on all equatorial states once it is fixed on $\ket{+}$.

Substituting \eqref{eq: condition for CC} into \eqref{eq: PCPIFC}, we obtain
\begin{align}\label{eq: aCC}
\Bcal(\ketbra{+}{+})=&\ (1-2c_4)(\ketbra{00}{00}+\ketbra{11}{11})+\left(2c_4-\frac{1}{2}\right)(\ketbra{01}{01}+\ketbra{10}{10})\nonumber\\
+&\ \frac{c_2}{2}(\ketbra{00}{11})(\bra{01}+\bra{10})+\frac{1-c_2}{2}(\ket{01}+\ket{10})(\bra{00}+\bra{11})\nonumber\\
+&\ \frac{1}{2}(\ketbra{01}{10}+\ketbra{10}{01}).
\end{align}
Taking partial traces yields
\begin{align}
\Tr_1[\Bcal(\ketbra{+}{+})]=\Tr_2[\Bcal(\ketbra{+}{+})]=\ketbra{+}{+}\;.
\end{align}
Therefore, by linearity and \texttt{PHASE}, $\Bcal$ satisfies $\Ssf'(\Hcal)$-\texttt{BROAD}. Moreover, since $\Tr_{12}[C_\Bcal]=\one$, $\Bcal$ also satisfies \texttt{TP} everywhere.

Hence, we obtain the following proposition.
\begin{prop}\label{prop: BC}
A linear map $\Bcal:\Lsf(\Hcal)\to\Lsf(\Hcal\otimes\Hcal)$ satisfying \texttt{\em PHASE}, \texttt{\em FLIP}, \texttt{\em PERM}, and \texttt{\em CLASSIC} also satisfies $\Ssf'(\Hcal)$-\texttt{\em BROAD} and \texttt{\em TP}.
\end{prop}

\begin{rem}
    Note that, if $\Bcal$ satisfies $\Ssf'(\Hcal)$-\texttt{\em BROAD}, it automatically follows that $\Bcal$ is \texttt{\em TP} for states of $\Ssf'(\Hcal)$. However, $\Tr_{12}[C_\Bcal]=\one$ means that $\Bcal$ is \texttt{\em TP} for the whole of $\Ssf(\Hcal)$. This does not follow, in principle, from $\Ssf'(\Hcal)$-\texttt{\em BROAD}  alone. The fact that $\Bcal$ is \texttt{\em TP} \emph{everywhere} is a non-trivial consequence of our analysis above. Note also that, at this point, \texttt{\em HP} need not be satisfied.
\end{rem}

\subsection{Singling out the optimal map}

Unlike the unitary covariant case, the operational properties introduced so far do not single out a unique map. We therefore introduce an optimization criterion to select, among all maps satisfying the required symmetries and consistency conditions, the one with minimal simulation cost. Our choice is thus to minimize $\|\Bcal\|_\blacklozenge$. Since the base norm is bounded in terms of the trace norm by~\eqref{eq: bound of simulation cost}, it is natural to search for parameters that minimize $\|C_\Bcal\|_1$.

Using the Wolfram Cloud platform~\cite{WolframCloud}, by numerically minimizing the trace norm of the Choi operator~\eqref{eq: PCPIFC} under the \texttt{CLASSIC} constraints~\eqref{eq: condition for CC} and the \texttt{HP} condition, which for $C_{\Bcal}$ is equivalent to $c_1,c_4\in\mathbb{R}$ and $c_2=c_3^*$, we obtain
\[
\min\|C_\Bcal\|_1=\frac{10}{3},
\]
with a unique minimizer given by
\[
c_1=\frac{1}{3},\quad
c_2=c_3=\frac{1}{2},\quad
c_4=\frac{5}{12},\quad
c_5=-\frac{1}{12},\quad
c_6=0.
\]
Using~\eqref{eq: bound of simulation cost}, every map $\Bcal$ satisfying \texttt{PHASE}, \texttt{PERM}, \texttt{FLIP}, \texttt{CLASSIC}, and \texttt{HP} must therefore obey
\begin{align}\label{eq: LB of base norm}
\|\Bcal\|_\blacklozenge\geq\frac{5}{3}.
\end{align}

The Choi operator achieving this minimum is
\begin{align}\label{eq: min one-norm}
\hat{C}_\Bcal\coloneq
\begin{bmatrix}
\frac{1}{3} & 0 & 0 & \frac{1}{2} & 0 & \frac{1}{2} & 0 & 0 \\
0 & 0 & 0 & 0 & 0 & 0 & 0 & 0 \\
0 & 0 & \frac{1}{3} & 0 & \frac{1}{2} & 0 & 0 & \frac{1}{2} \\
\frac{1}{2} & 0 & 0 & \frac{1}{3} & 0 & \frac{1}{2} & 0 & 0 \\
0 & 0 & \frac{1}{2} & 0 & \frac{1}{3} & 0 & 0 & \frac{1}{2} \\
\frac{1}{2} & 0 & 0 & \frac{1}{2} & 0 & \frac{1}{3} & 0 & 0 \\
0 & 0 & 0 & 0 & 0 & 0 & 0 & 0 \\
0 & 0 & \frac{1}{2} & 0 & \frac{1}{2} & 0 & 0 & \frac{1}{3} \\
\end{bmatrix}.
\end{align}

This operator admits a decomposition into two positive semidefinite operators with orthogonal support:
\begin{align}
\hat{C}_\Bcal=\frac{4}{3}\hat{C}_\Bcal^+-\frac{1}{3}\hat{C}_\Bcal^-\;,
\end{align}
where the positive and negative parts are
\begin{align}
\hat{C}_\Bcal^+&\coloneq
\begin{bmatrix}\label{eq: positive part}
 \frac{1}{3} & 0 & 0 & \frac{1}{3} & 0 & \frac{1}{3} & 0 & 0 \\
   0 & 0 & 0 & 0 & 0 & 0 & 0 & 0 \\
   0 & 0 & \frac{1}{3} & 0 & \frac{1}{3} & 0 & 0 & \frac{1}{3} \\
   \frac{1}{3} & 0 & 0 & \frac{1}{3} & 0 & \frac{1}{3} & 0 & 0 \\
   0 & 0 & \frac{1}{3} & 0 & \frac{1}{3} & 0 & 0 & \frac{1}{3} \\
   \frac{1}{3} & 0 & 0 & \frac{1}{3} & 0 & \frac{1}{3} & 0 & 0 \\
   0 & 0 & 0 & 0 & 0 & 0 & 0 & 0 \\
   0 & 0 & \frac{1}{3} & 0 & \frac{1}{3} & 0 & 0 & \frac{1}{3} \\
\end{bmatrix}
\end{align}
and
\begin{align}
\hat{C}_\Bcal^-&\coloneq
\begin{bmatrix}\label{eq: negative part}
 \frac{1}{3} & 0 & 0 & -\frac{1}{6} & 0 & -\frac{1}{6} & 0 & 0 \\
   0 & 0 & 0 & 0 & 0 & 0 & 0 & 0 \\
   0 & 0 & \frac{1}{3} & 0 & -\frac{1}{6} & 0 & 0 & -\frac{1}{6} \\
   -\frac{1}{6} & 0 & 0 & \frac{1}{3} & 0 & -\frac{1}{6} & 0 & 0 \\
   0 & 0 & -\frac{1}{6} & 0 & \frac{1}{3} & 0 & 0 & -\frac{1}{6} \\
   -\frac{1}{6} & 0 & 0 & -\frac{1}{6} & 0 & \frac{1}{3} & 0 & 0 \\
   0 & 0 & 0 & 0 & 0 & 0 & 0 & 0 \\
   0 & 0 & -\frac{1}{6} & 0 & -\frac{1}{6} & 0 & 0 & \frac{1}{3} \\
\end{bmatrix}.
\end{align}
A direct calculation shows that both operators correspond to \texttt{CPTP} maps:
\begin{align}
    \Tr_{1,2}[\hat{C}_\Bcal^+]=\Tr_{1,2}[\hat{C}_\Bcal^-]=\one\;.
\end{align}

As in the unitary covariant case~\cite{buscemi2013direct,parzygnat2024virtual}, this decomposition allows us to compute the simulation cost immediately:
\begin{prop}
For any linear map $\Bcal:\Lsf(\Hcal)\to\Lsf(\Hcal\otimes\Hcal)$ satisfying \texttt{\em PHASE}, \texttt{\em FLIP}, \texttt{\em PERM}, \texttt{\em CLASSIC}, and \texttt{\em HP}, it holds that $\|\Bcal\|_\blacklozenge=\frac{5}{3}$.
\end{prop}

In the unitary covariant case~\cite{xiao2025noPRL, xiao2025noPRR}, the simulation cost of the virtual broadcasting map is $d=\dim(\Hcal)$. Thus, for qubits, relaxing unitary covariance to phase-covariance indeed lowers the simulation cost. However,
\begin{align}
\left(\frac{5}{3}\right)^2 n_Q > 2 n_Q \geq n_1+n_2,
\end{align}
so \texttt{SAMPLE} is still not satisfied. We therefore obtain:
\begin{cor}
No linear map exists that simultaneously satisfies \texttt{\em PHASE}, \texttt{\em FLIP}, \texttt{\em PERM}, \texttt{\em CLASSIC}, \texttt{\em HP}, and \texttt{\em SAMPLE}.
\end{cor}

Notably, $\hat{C}_\Bcal^+$ coincides with the Choi operator of the optimal phase-covariant cloning map derived in~\cite{d2003optimal}. This yields the following proposition.
\begin{prop}
Let $\hat{\Bcal}$ be the \texttt{\em HPTP} map corresponding to $\hat{C}_\Bcal$ in Eq.~\eqref{eq: min one-norm}. Then, it holds that
\begin{align}
    \min_{\Ecal:\text{\texttt{\em CPTP}}}\|\hat{\Bcal}-\Ecal\|_\lozenge=\frac{2}{3}\;,
\end{align}
where $\|\Ecal\|_\lozenge\coloneq\max_{\rho\in\Ssf(\Hcal\otimes\Hcal)}\|(\Ecal\otimes\Ical)(\rho)\|_1$ is the diamond norm. The minimum is achieved when $\Ecal$ is the optimal phase-covariant cloning map.
\end{prop}

The argument follows the same line as in~\cite{parzygnat2024virtual}. This result is the phase-covariant analogue of the fact established in~\cite{parzygnat2024virtual}: the \texttt{CPTP} map closest to the canonical broadcasting map is the optimal unitary covariant cloning map. Moreover, for the canonical case, $\|\Bcal-\Ecal\|_\lozenge=d-1$. Hence, for qubits, relaxing unitary covariance to phase-covariance makes the virtual broadcasting map strictly closer to a \texttt{CPTP} map.

\section{Conclusion}\label{sec: Conclusion}

In this work, we studied virtual phase-covariant quantum broadcasting for qubits. By imposing phase-covariance, flip covariance, permutation invariance, and classical consistency, we showed that the structure of a virtual broadcasting map is strongly constrained and can be fully characterized. Within this class, we identified the unique map that minimizes the simulation cost. We also demonstrated that, in the phase-covariant setting, both the simulation cost and the distance to the closest physical map are strictly smaller than in the unitary covariant case. Furthermore, we established that, as in the unitary covariant scenario, the physical map closest to the virtual broadcasting map is the corresponding phase-covariant cloning map. Despite these improvements, we proved that the resulting virtual broadcasting map remains impractical, as it still fails to satisfy the sample-efficiency criterion.

Our analysis focused on qubits. A natural next step is to extend this framework to qudits. Earlier results~\cite{parzygnat2024virtual,yao2024optimal,xiao2025noPRL,xiao2025noPRR,yao2025quantifying} rely on tools from representation theory and semidefinite optimization that exploit full unitary covariance. Since our assumptions are weaker, such methods do not apply directly, but related ideas may still be adapted effectively.

Another promising research direction is to increase the number of input copies and study the redistribution \textit{rate}. In the unitary covariant case, and also in the phase-covariant case, it is known that the rate of perfect broadcasting becomes infinite already with a \texttt{CPTP} linear map once six input copies (unitary covariant case) or four input copies (phase-covariant case) are available~\cite{buscemi2006universal}. It is plausible that, for virtual maps, this threshold can be lowered even further. Understanding whether virtual operations allow perfect or near-perfect redistribution at smaller copy numbers may reveal new structural features that do not appear in the physical setting.

From an operational perspective, an important open question is to characterize virtual broadcasting maps that satisfy sample efficiency. Even after relaxing unitary covariance to phase-covariance, this requirement remained out of reach. Understanding how operational usefulness constrains virtual operations may shed light on the limits of simulating quantum processes and the boundary between physical and virtual transformations.

\section*{Acknowledgements} 
The authors acknowledge support from MEXT Quantum Leap Flagship Program (MEXT QLEAP) Grant No. JPMXS0120319794; from MEXT-JSPS  Grant-in-Aid for Transformative Research Areas  (A) ``Extreme Universe,'' No.~21H05183; and  from JSPS  KAKENHI Grant No.~23K03230.

\bibliographystyle{unsrt} 
\bibliography{main.bib}

\begin{thebibliography}{10}

\bibitem{wootters1982single}
William~K Wootters and Wojciech~H Zurek.
\newblock A single quantum cannot be cloned.
\newblock {\em Nature}, 299(5886):802--803, 1982.

\bibitem{DIEKS1982271}
D.~Dieks.
\newblock Communication by epr devices.
\newblock {\em Physics Letters A}, 92(6):271--272, 1982.

\bibitem{barnum1996noncommuting}
Howard Barnum, Carlton~M Caves, Christopher~A Fuchs, Richard Jozsa, and Benjamin Schumacher.
\newblock Noncommuting mixed states cannot be broadcast.
\newblock {\em Physical Review Letters}, 76(15):2818, 1996.

\bibitem{parzygnat2024virtual}
Arthur~J Parzygnat, James Fullwood, Francesco Buscemi, and Giulio Chiribella.
\newblock Virtual quantum broadcasting.
\newblock {\em Physical Review Letters}, 132(11):110203, 2024.

\bibitem{yao2024optimal}
Hongshun Yao, Xia Liu, Chengkai Zhu, and Xin Wang.
\newblock Optimal unilocal virtual quantum broadcasting.
\newblock {\em Physical Review A}, 110(1):012458, 2024.

\bibitem{zheng2025experimental}
Yuxuan Zheng, Xinfang Nie, Hongfeng Liu, Yutong Luo, Dawei Lu, and Xiangjing Liu.
\newblock Experimental virtual quantum broadcasting.
\newblock {\em Physical Review A}, 111(6):L060402, 2025.

\bibitem{xiao2025noPRL}
Yunlong Xiao, Xiangjing Liu, and Zhenhuan Liu.
\newblock No practical quantum broadcasting: Even virtually.
\newblock {\em Physical Review Letters}, 135(9):090202, 2025.

\bibitem{xiao2025noPRR}
Yunlong Xiao, Xiangjing Liu, and Zhenhuan Liu.
\newblock No practical quantum broadcasting: General framework.
\newblock {\em Physical Review Research}, 7(3):033194, 2025.

\bibitem{yao2025quantifying}
Hongshun Yao, Jingu Xie, Xuanqiang Zhao, Chengkai Zhu, Ranyiliu Chen, and Xin Wang.
\newblock Quantifying unxtendibility via virtual state extension.
\newblock {\em arXiv preprint arXiv:2510.24895}, 2025.

\bibitem{buscemi2013direct}
Francesco Buscemi, Michele Dall'Arno, Masanao Ozawa, and Vlatko Vedral.
\newblock Direct observation of any two-point quantum correlation function.
\newblock {\em arXiv preprint arXiv:1312.4240}, 2013.

\bibitem{buscemi2014universal}
Francesco Buscemi, Michele Dall'Arno, Masanao Ozawa, and Vlatko Vedral.
\newblock Universal optimal quantum correlator.
\newblock {\em International Journal of Quantum Information}, 12(07n08):1560002, 2014.

\bibitem{rossini2023single}
Mirko Rossini, Dominik Maile, Joachim Ankerhold, and Brecht~IC Donvil.
\newblock Single-qubit error mitigation by simulating non-markovian dynamics.
\newblock {\em Physical Review Letters}, 131(11):110603, 2023.

\bibitem{temme2017error}
Kristan Temme, Sergey Bravyi, and Jay~M Gambetta.
\newblock Error mitigation for short-depth quantum circuits.
\newblock {\em Physical review letters}, 119(18):180509, 2017.

\bibitem{jiang2021physical}
Jiaqing Jiang, Kun Wang, and Xin Wang.
\newblock Physical implementability of linear maps and its application in error mitigation.
\newblock {\em Quantum}, 5:600, 2021.

\bibitem{takagi2022fundamental}
Ryuji Takagi, Suguru Endo, Shintaro Minagawa, and Mile Gu.
\newblock Fundamental limits of quantum error mitigation.
\newblock {\em npj Quantum Information}, 8(1):114, 2022.

\bibitem{fullwood2022quantum}
James Fullwood and Arthur~J Parzygnat.
\newblock On quantum states over time.
\newblock {\em Proceedings of the Royal Society A}, 478(2264):20220104, 2022.

\bibitem{parzygnat2023time}
Arthur~J Parzygnat and James Fullwood.
\newblock From time-reversal symmetry to quantum bayes’ rules.
\newblock {\em PRX Quantum}, 4(2):020334, 2023.

\bibitem{lie2023uniqueness}
Seok~Hyung Lie and Nelly~HY Ng.
\newblock Uniqueness of quantum state over time function.
\newblock {\em arXiv preprint arXiv:2308.12752}, 2023.

\bibitem{werner1998optimal}
Reinhard~F Werner.
\newblock Optimal cloning of pure states.
\newblock {\em Physical Review A}, 58(3):1827, 1998.

\bibitem{d2001optimal}
G~Mauro D’Ariano and P~Lo Presti.
\newblock Optimal nonuniversally covariant cloning.
\newblock {\em Physical Review A}, 64(4):042308, 2001.

\bibitem{d2003optimal}
Giacomo~Mauro D’Ariano and Chiara Macchiavello.
\newblock Optimal phase-covariant cloning for qubits and qutrits.
\newblock {\em Physical Review A}, 67(4):042306, 2003.

\bibitem{buscemi2005economical}
Francesco Buscemi, Giacomo~Mauro D’Ariano, and Chiara Macchiavello.
\newblock Economical phase-covariant cloning of qudits.
\newblock {\em Physical Review A―Atomic, Molecular, and Optical Physics}, 71(4):042327, 2005.

\bibitem{regula2021operational}
Bartosz Regula, Ryuji Takagi, and Mile Gu.
\newblock Operational applications of the diamond norm and related measures in quantifying the non-physicality of quantum maps.
\newblock {\em Quantum}, 5:522, 2021.

\bibitem{WolframCloud}
Inc. Wolfram~Research.
\newblock Wolfram cloud, 2025.
\newblock Available at \url{https://www.wolframcloud.com}.

\bibitem{buscemi2006universal}
Francesco Buscemi, Giacomo~Mauro D’Ariano, Chiara Macchiavello, and Paolo Perinotti.
\newblock Universal and phase-covariant superbroadcasting for mixed qubit states.
\newblock {\em Physical Review A―Atomic, Molecular, and Optical Physics}, 74(4):042309, 2006.

\end{thebibliography}
\end{document}